\documentclass[%
 aip,
 amsmath,
 amssymb, 
reprint,%
]{revtex4-1}
\usepackage{graphicx}
\usepackage{dcolumn}
\usepackage{epstopdf} 
\usepackage{color}
\usepackage[mathlines]{lineno}
 \makeatletter
  \providecommand\color[2][]{%
    \GenericError{(gnuplot) \space\space\space\@spaces}{%
      Package color not loaded in conjunction with
      terminal option `colourtext'%
    }{See the gnuplot documentation for explanation.%
    }{Either use 'blacktext' in gnuplot or load the package
      color.sty in LaTeX.}%
    \renewcommand\color[2][]{}%
  }%
  \providecommand\includegraphics[2][]{%
    \GenericError{(gnuplot) \space\space\space\@spaces}{%
      Package graphicx or graphics not loaded%
    }{See the gnuplot documentation for explanation.%
    }{The gnuplot epslatex terminal needs graphicx.sty or graphics.sty.}%
    \renewcommand\includegraphics[2][]{}%
  }%
  \providecommand\rotatebox[2]{#2}%
  \@ifundefined{ifGPcolor}{%
    \newif\ifGPcolor
    \GPcolorfalse
  }{}%
  \@ifundefined{ifGPblacktext}{%
    \newif\ifGPblacktext
    \GPblacktexttrue
  }{}%
  \let\gplgaddtomacro\g@addto@macro

  \makeatother
 
\begin{document}

\title{ Fermi level pinning at the  Ge(001) surface --- A case for non-standard explanation}

\author{Mateusz Wojtaszek}
\affiliation{Center for Nanometer-Scale Science and Advanced Materials (NANOSAM), Faculty of Physics, Astronomy and Applied Computer Science, Jagiellonian University, ul. St. Lojasiewicza 11, 30-348 Krakow, Poland}
\author{Rafal Zuzak}
\affiliation{Center for Nanometer-Scale Science and Advanced Materials (NANOSAM), Faculty of Physics, Astronomy and Applied Computer Science, Jagiellonian University, ul. St. Lojasiewicza 11, 30-348 Krakow, Poland}
\author{Szymon Godlewski}
\affiliation{Center for Nanometer-Scale Science and Advanced Materials (NANOSAM), Faculty of Physics, Astronomy and Applied Computer Science, Jagiellonian University, ul. St. Lojasiewicza 11, 30-348 Krakow, Poland}
\author{Marek Kolmer}
\affiliation{Center for Nanometer-Scale Science and Advanced Materials (NANOSAM), Faculty of Physics, Astronomy and Applied Computer Science, Jagiellonian University, ul. St. Lojasiewicza 11, 30-348 Krakow, Poland}

\author{Jakub Lis}
\email{j.lis@uj.edu.pl}
\affiliation{Center for Nanometer-Scale Science and Advanced Materials (NANOSAM), Faculty of Physics, Astronomy and Applied Computer Science, Jagiellonian University, ul. St. Lojasiewicza 11, 30-348 Krakow, Poland}

\author{Bartosz Such}
\affiliation{Center for Nanometer-Scale Science and Advanced Materials (NANOSAM), Faculty of Physics, Astronomy and Applied Computer Science, Jagiellonian University, ul. St. Lojasiewicza 11, 30-348 Krakow, Poland}

\author{Marek Szymonski}
\affiliation{Center for Nanometer-Scale Science and Advanced Materials (NANOSAM), Faculty of Physics, Astronomy and Applied Computer Science, Jagiellonian University, ul. St. Lojasiewicza 11, 30-348 Krakow, Poland}

\pacs{}
\keywords{hydrogenated semiconductor, germanium, band bending, four point probe, scanning tunneling microscopy}
\date{\today}

\begin{abstract}      
To explore the origin of the Fermi level pinning in germanium we investigate the Ge(001) and Ge(001):H surfaces. The absence of  relevant surface states in the case of Ge(001):H  should unpin the surface Fermi level. This  is not observed. For samples with donors as majority dopants  the surface Fermi level appears close to the top of the valence band regardless of the surface structure. Surprisingly, for the passivated surface it is  located  below the top of the valence band allowing  scanning tunneling microscopy imaging within the band gap.  We argue  that the well known electronic mechanism behind band bending does not apply and a more complicated scenario  involving ionic degrees of freedom is  therefore  necessary.  Experimental techniques involve four point probe electric current measurements, scanning tunneling microscopy and  spectroscopy. 
\end{abstract}     
\maketitle
\section{Introduction}
The renewed interest in germanium is driven mostly by  the prospects it offers to microelectronics \cite{Avenues,Bracht}. To this end a detailed understanding of surface electronic and structural properties is essential. The position of the  surface Fermi level (SFL) and the structure of surface states are key features in this context. 

There is a generally accepted theory of the Fermi level pinning (FLP)  in semiconductors,  formulated decades ago~\cite{Monch}, which operates within the electronic degrees of freedom only. 
This understanding is useful in explaining many phenomena, e.g independence of the SFL from the bulk doping and creation of two-dimensional electron gases~\cite{InAs}. For germanium surfaces this model was adapted by P.~Tsipas \emph{et al.} in Ref.~\onlinecite{Greeks}. Nevertheless, the strength and nature of FLP in germanium is still under debate as outlined recently~\cite{Demkov}.

In our recent paper~\cite{APL} we demonstrated that the surface Fermi level is pinned close to the valence band maximum for  Ge(001). In that letter we concentrated on the conductance data and their interpretation. We  suggested  that the SFL is pinned by the surface states (dangling bonds) due to the  fact that there are no additional structures on the surface capable  of pinning the SFL.  Our conductivity experiments reported below are consistent with the surface Fermi position  revealed by Scanning Tunneling Spectroscopy (STS). Surprisingly, these data suggest a metallic character for the Ge(001):H surface resulting from the SFL pinned below the top of the valence band. 
What we find of particular  interest  is that the surface states recovered from STS seem to  be at odds with the well-known mechanism behind FLP. We also review experimental reports, mostly based on angle resolved photoelectron spectroscopy (ARPES), suggesting  a  failure of the electronic theory of the FLP for germanium. We postulate that a mechanism similar to that valid for silicon is in effect.
\\

\section{Experimental}
The experiments were performed on three different germanium (MTI Corporation) samples: undoped ($\sim 45\ \Omega\ cm$), n-type Sb-doped ($0.008-0.009\ \Omega\ cm$) and p-type Ga-doped ($0.1-0.5\ \Omega\ cm$). The intrinsic doping for germanium is n-type, with the density of carriers about~$10^{13} \ cm^{-3}$. The experiments were done  in an ultra-high vacuum (UHV) system with a base pressure below $5\cdot 10^{-10}$ mbar. The sample preparation consisted of subsequent sputtering ($Ar^{+}$, $800 \ eV$) and annealing  to $1040 \ K$ cycles. The cleanness of the (001) surface was checked with the low energy electron diffraction and the scanning tunneling microscopy (STM).  Afterwards, to passivate the surface, the substrate  was exposed to atomic hydrogen flux dosed with a home-built hydrogen cracker ($10^{-7}$~mbar partial pressure, sample temperature 480~K). The procedure  resulted  in a well-hydrogenated surface with the defect density down to 2-4\% of surface dimers, which translates to a defect density up to  $4\cdot 10^{12}$ cm$^{-2}$. Further details of the preparation process may be found in~Ref.~\onlinecite{Kolmer}. Prepared samples were transported in the UHV conditions to the surface transport measurement chamber or to the STM chamber.
The surface transport measurements were carried out at room temperature in UHV Nanoprobe System (Omicron Nanotechnology GmbH) equipped with  a high resolution scanning electron microscope  (Gemini Column, Carl Zeiss). Scanning tunneling spectroscopy and microscopy experiments were done at the liquid helium temperature (4.5~K). \\

\section{Experiments and Discussion}
\subsection{Resistance analysis}
\begin{figure}
\resizebox{!}{0.32\textwidth}{
\begingroup
  \gdef\gplbacktext{}%
  \gdef\gplfronttext{}%
  \ifGPblacktext
    \def\colorrgb#1{}%
    \def\colorgray#1{}%
  \else
    \ifGPcolor
      \def\colorrgb#1{\color[rgb]{#1}}%
      \def\colorgray#1{\color[gray]{#1}}%
      \expandafter\def\csname LTw\endcsname{\color{white}}%
      \expandafter\def\csname LTb\endcsname{\color{black}}%
      \expandafter\def\csname LTa\endcsname{\color{black}}%
      \expandafter\def\csname LT0\endcsname{\color[rgb]{1,0,0}}%
      \expandafter\def\csname LT1\endcsname{\color[rgb]{0,1,0}}%
      \expandafter\def\csname LT2\endcsname{\color[rgb]{0,0,1}}%
      \expandafter\def\csname LT3\endcsname{\color[rgb]{1,0,1}}%
      \expandafter\def\csname LT4\endcsname{\color[rgb]{0,1,1}}%
      \expandafter\def\csname LT5\endcsname{\color[rgb]{1,1,0}}%
      \expandafter\def\csname LT6\endcsname{\color[rgb]{0,0,0}}%
      \expandafter\def\csname LT7\endcsname{\color[rgb]{1,0.3,0}}%
      \expandafter\def\csname LT8\endcsname{\color[rgb]{0.5,0.5,0.5}}%
    \else
      \def\colorrgb#1{\color{black}}%
      \def\colorgray#1{\color[gray]{#1}}%
      \expandafter\def\csname LTw\endcsname{\color{white}}%
      \expandafter\def\csname LTb\endcsname{\color{black}}%
      \expandafter\def\csname LTa\endcsname{\color{black}}%
      \expandafter\def\csname LT0\endcsname{\color{black}}%
      \expandafter\def\csname LT1\endcsname{\color{black}}%
      \expandafter\def\csname LT2\endcsname{\color{black}}%
      \expandafter\def\csname LT3\endcsname{\color{black}}%
      \expandafter\def\csname LT4\endcsname{\color{black}}%
      \expandafter\def\csname LT5\endcsname{\color{black}}%
      \expandafter\def\csname LT6\endcsname{\color{black}}%
      \expandafter\def\csname LT7\endcsname{\color{black}}%
      \expandafter\def\csname LT8\endcsname{\color{black}}%
    \fi
  \fi
  \setlength{\unitlength}{0.0500bp}%
  \begin{picture}(7200.00,5040.00)%
    \gplgaddtomacro\gplbacktext{%
      \csname LTb\endcsname%
      \put(1078,704){\makebox(0,0)[r]{\strut{} 0}}%
      \put(1078,1673){\makebox(0,0)[r]{\strut{} 1000}}%
      \put(1078,2643){\makebox(0,0)[r]{\strut{} 2000}}%
      \put(1078,3612){\makebox(0,0)[r]{\strut{} 3000}}%
      \put(1078,4581){\makebox(0,0)[r]{\strut{} 4000}}%
      \put(1210,484){\makebox(0,0){\strut{} 0}}%
      \put(2542,484){\makebox(0,0){\strut{} 0.25}}%
      \put(3873,484){\makebox(0,0){\strut{} 0.5}}%
      \put(5205,484){\makebox(0,0){\strut{} 0.75}}%
      \put(6537,484){\makebox(0,0){\strut{} 1}}%
      \put(176,2739){\rotatebox{-270}{\makebox(0,0){R [$\Omega$]}}}%
      \put(4006,154){\makebox(0,0){\large{$x$}}}%
    }%
    \gplgaddtomacro\gplfronttext{%
      \csname LTb\endcsname%
      \put(2530,4602){\makebox(0,0)[r]{\strut{}Ge(001)}}%
      \csname LTb\endcsname%
      \put(2530,4382){\makebox(0,0)[r]{\strut{}Ge(001):H}}%
    }%
    \gplbacktext
    \put(0,0){\includegraphics{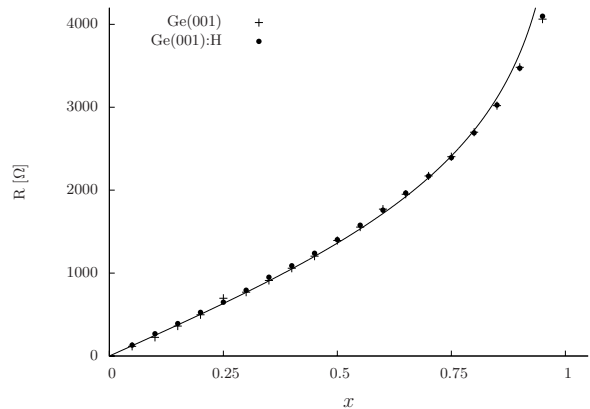}}%
    \gplfronttext
  \end{picture}%
\endgroup
}
\caption{\label{Ntype} Data acquired on n-type doped sample for pure Ge(001) and  after surface passivation for the distance $D=8 \ \mu$m between the current source and drain. The distance $x$ on the horizontal axis shows the distance between electrodes measuring the voltage drop as a fraction of $D$, see Ref.~\onlinecite{APL}. For different surfaces, the  measurement outcomes do not differ much. Fitting to formula~(\ref{2dim}) gives two-dimensional  conductivity $\sigma_{2}=(2.61\pm 0.03) \cdot 10^{-4} \square /  \Omega$ for the bare surface and $\sigma_{2}=(2.57\pm 0.02) \cdot 10^{-4} \square /  \Omega$,  which is close to the values reported in Ref.~\onlinecite{APL}. The smooth curve shows the function~(\ref{2dim}) with the parameter $\sigma_{2}$ fitted to the Ge(001):H data.}
\end{figure}

\begin{figure}
\resizebox{!}{0.32\textwidth}{
\begingroup
 \gdef\gplbacktext{}%
  \gdef\gplfronttext{}%
  \ifGPblacktext
    \def\colorrgb#1{}%
    \def\colorgray#1{}%
  \else
    \ifGPcolor
      \def\colorrgb#1{\color[rgb]{#1}}%
      \def\colorgray#1{\color[gray]{#1}}%
      \expandafter\def\csname LTw\endcsname{\color{white}}%
      \expandafter\def\csname LTb\endcsname{\color{black}}%
      \expandafter\def\csname LTa\endcsname{\color{black}}%
      \expandafter\def\csname LT0\endcsname{\color[rgb]{1,0,0}}%
      \expandafter\def\csname LT1\endcsname{\color[rgb]{0,1,0}}%
      \expandafter\def\csname LT2\endcsname{\color[rgb]{0,0,1}}%
      \expandafter\def\csname LT3\endcsname{\color[rgb]{1,0,1}}%
      \expandafter\def\csname LT4\endcsname{\color[rgb]{0,1,1}}%
      \expandafter\def\csname LT5\endcsname{\color[rgb]{1,1,0}}%
      \expandafter\def\csname LT6\endcsname{\color[rgb]{0,0,0}}%
      \expandafter\def\csname LT7\endcsname{\color[rgb]{1,0.3,0}}%
      \expandafter\def\csname LT8\endcsname{\color[rgb]{0.5,0.5,0.5}}%
    \else
      \def\colorrgb#1{\color{black}}%
      \def\colorgray#1{\color[gray]{#1}}%
      \expandafter\def\csname LTw\endcsname{\color{white}}%
      \expandafter\def\csname LTb\endcsname{\color{black}}%
      \expandafter\def\csname LTa\endcsname{\color{black}}%
      \expandafter\def\csname LT0\endcsname{\color{black}}%
      \expandafter\def\csname LT1\endcsname{\color{black}}%
      \expandafter\def\csname LT2\endcsname{\color{black}}%
      \expandafter\def\csname LT3\endcsname{\color{black}}%
      \expandafter\def\csname LT4\endcsname{\color{black}}%
      \expandafter\def\csname LT5\endcsname{\color{black}}%
      \expandafter\def\csname LT6\endcsname{\color{black}}%
      \expandafter\def\csname LT7\endcsname{\color{black}}%
      \expandafter\def\csname LT8\endcsname{\color{black}}%
    \fi
  \fi
  \setlength{\unitlength}{0.0500bp}%
  \begin{picture}(7200.00,5040.00)%
    \gplgaddtomacro\gplbacktext{%
      \csname LTb\endcsname%
      \put(1078,704){\makebox(0,0)[r]{\strut{} 0}}%
      \put(1078,1576){\makebox(0,0)[r]{\strut{} 300}}%
      \put(1078,2449){\makebox(0,0)[r]{\strut{} 600}}%
      \put(1078,3321){\makebox(0,0)[r]{\strut{} 900}}%
      \put(1078,4193){\makebox(0,0)[r]{\strut{} 1200}}%
      \put(1210,484){\makebox(0,0){\strut{} 0}}%
      \put(2542,484){\makebox(0,0){\strut{} 0.25}}%
      \put(3873,484){\makebox(0,0){\strut{} 0.5}}%
      \put(5205,484){\makebox(0,0){\strut{} 0.75}}%
      \put(6537,484){\makebox(0,0){\strut{} 1}}%
      \put(176,2739){\rotatebox{-270}{\makebox(0,0){\strut{}R [$\Omega$]}}}%
      \put(4006,154){\makebox(0,0){\large{$x$}}}%
    }%
    \gplgaddtomacro\gplfronttext{%
      \csname LTb\endcsname%
      \put(2530,4602){\makebox(0,0)[r]{\strut{}Ge(001)}}%
      \csname LTb\endcsname%
      \put(2530,4382){\makebox(0,0)[r]{\strut{}Ge(001):H}}%
    }%
    \gplbacktext
    \put(0,0){\includegraphics{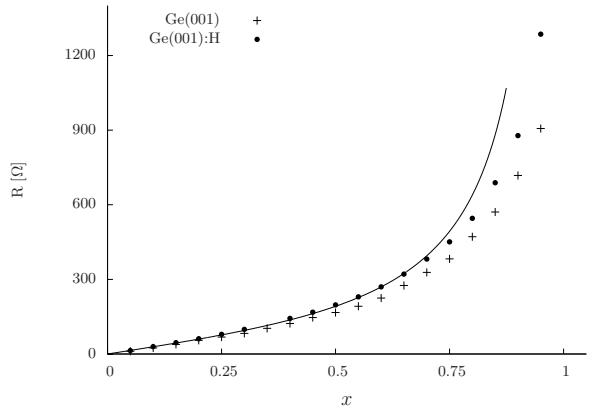}}%
    \gplfronttext
  \end{picture}%
\endgroup
}
    \caption{\label{Ptype} Representative data acquired for p-type doped sample for pure Ge(001) and passivated surface for $D=8 \ \mu$m.   For different surfaces, the  measurement outcomes are similar, lower resistance for Ge(001) than for Ge(001):H is not systematically observed. Here, fitting results in $\sigma_{3}=(1.62\pm 0.01)\Omega^{-1}cm^{-1}$ for the bare surface and $\sigma_{3}=(1.38\pm 0.01)  \Omega^{-1}cm^{-1}$ for the hydrogenated one. The smooth curve shows  relation~(\ref{3dim}) with the parameter fitted for Ge(001):H. Poor fitting quality for $x\sim 1$ can be explained~\cite{PRB} as a result of conductivity variation near the surface.}

\end{figure}

In the previous paper~\cite{APL} we  reported resistance measurements on clean Ge(001) for three samples: n-type, p-type and intrinsically doped. The results were divided into two  qualitatively different groups: for n-type samples two dimensional current flow was observed while for p-type samples a  three dimensional current character was seen. For the explored distances between the electrodes the current  confinement in a thin layer beneath the surface is not a trivial effect  resulting from finite sample width as its occurrence correlates with the doping type. The reduced dimensionality  of the current was interpreted as a result of  FLP followed by  the  creation of an inversion layer beneath the surfaces on n-type doped samples.\\ 
In that paper~\cite{APL} we discussed the dimensional assignment in more details. Starting from the current flow equation, we showed  that current confinement in the direction normal to the surface results in a scale independence of resistance measurements. In the case of three-dimensional currents the results depend on the absolute distances between electrodes. The experimental points closely reproduced the relevant scaling laws~\cite{APL}.\\  
Here  we report on resistance  measurements  for Ge(001):H  in the same  experimental settings as in Ref.~\onlinecite{APL}. We applied the four point probe method. The current source and  drain were placed at distance $D$ (between 2 and 16 $\mu$m) while the distance between two probes measuring the voltage drop was changed. All measurements were done in a colinear geometry symmetric with the midpoint of the segment set by the current supplementing electrodes. In these settings the formula for measured resistance in the case of two-dimensional current flow yields
\begin{equation}\label{2dim}
R=\frac{1}{\pi\sigma_{2}}\ln{\frac{1+x}{1-x}},
\end{equation}
where $\sigma_{2}$ is the two-dimensional conductance and $x$ is the distance between the probing electrodes normalized with $D$. For the three-dimensional currents the formula reads
\begin{equation}\label{3dim}
R=\frac{1}{\pi D \sigma_{3}}\frac{x}{1-x^{2}},
\end{equation}
where $\sigma_{3}$ stands for the three-dimensional conductivity. A more detailed analysis of the above relation may be found in Ref.~\onlinecite{PRB}. These formulas highlight the qualitative difference between the two-dimensional and three-dimensional current flows, suffice it to say that  $\sigma_{2}$ and $\sigma_{3}$ are expressed using different units. \\
 Fig.~\ref{Ntype} and \ref{Ptype} show representative data obtained for the Ge(001) and Ge(001):H surfaces for n-type and p-type doped samples (similar agreement is observed for the intrinsic sample). There are small variations of the conductance introduced by surface treatment. These differences are within conductivity fluctuations  observed for samples cut from one wafer. Consequently, the dimensional character  of the current flow is not altered.  This  provides a hint that the SFL is insensitive to the  actual  structure of surface states. In other words, it suggests that the inversion layer stays intact upon hydrogen termination of the surface.
  
\subsection{Scanning tunneling spectroscopy and microscopy data}
\begin{figure}
\includegraphics[scale=0.5]{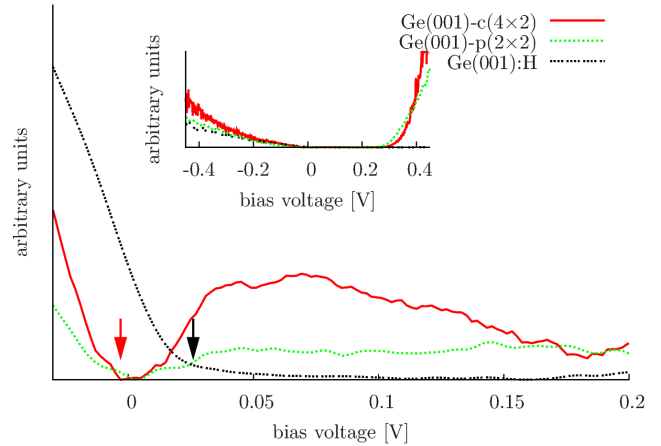}
    \caption{\label{STS} Spectroscopic data for Ge(001)-p(2$\times$2), Ge(001)-c(4$\times$2) and Ge(001):H acquired at 4.5~K, the feedback set-point in all cases was -0.5~V and~500 pA. The arrows mark the onset of the valence band.  Note  that the signal is extremely weak, unnoticeable when bands are observed. The data  in a broader range showing the surface band gap for the passivated surface can be found in Ref.~\onlinecite{Kolmer}. The quick change of $dI/dV$ visible in the inset for the bare surfaces for about 0.4~V   starts about 0.2~V in the close view.  }
\end{figure}

\begin{figure}
   \includegraphics[width=0.4\textwidth]{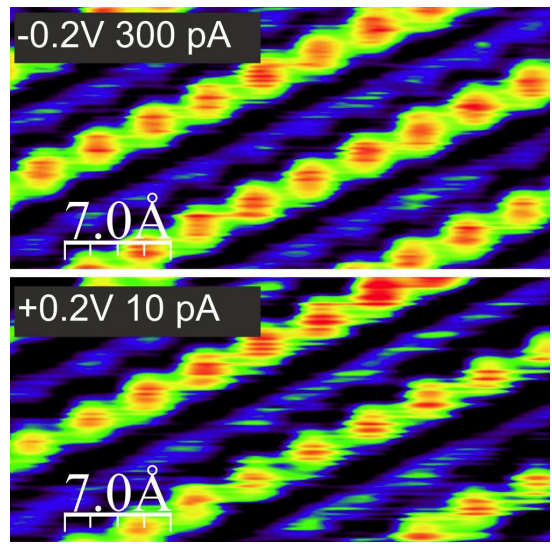}
    \caption{\label{STM}STM images of Ge(001):H acquired for -0.2~V bias voltage and +0.2 bias voltage, upper and lower panel respectively. The nearly indiscernible contrast appears as a result of imaging the valence band states above the SFL.}
\end{figure}

We concentrate on data acquired for the intrinsically doped sample. As it is effectively n-type doped, the bulk Fermi level is located above the mid-gap. For n-type doped sample the results are very similar. The case of the p-type doped sample is not that interesting, as the band bending is less obvious. 
The STS data obtained for bare Ge(001) and hydrogen passivated surfaces are shown in Fig.~\ref{STS}. In both cases there is a plateau corresponding to the surface band gap. It is about 0.2~V wide for Ge(001) and 0.85~V wide for Ge(001):H, see inset in Fig.~\ref{STS} and Ref.~\onlinecite{Kolmer}. For the passivated surface  $dI/dV$ for the zero bias voltage does not vanish. It suggests  that the SFL is located below the top of the valence band. Indeed, we can repetitively acquire STM images for any positive bias voltage  at the Ge(001):H surface, see Fig.~\ref{STM}. The resulting pattern coincides with the contrast obtained at negative biases. This makes us  believe  that the SFL in this case is located slightly (10-30 meV) below the top of the valence band. \\
In the case of  Ge(001), the SFL is also located at the top of the valence band, as suggested in Fig. 3. We cannot obtain repetitively STM images for small positive biases (up to 0.2~V). However, on rare occasions we happen to arrive at  tip apexes  that admit such STM imaging. This  suggests  that  STS and STM experiments allow determination of the SFL within several tens of milivolts accuracy. Such  indeterminacy appears to be an intrinsic feature of the method linked to tip-induced band bending~\cite{repp}. Furthermore, the bare surface is much more chemically reactive than its passivated counterpart. As such, the tip probably cannot  approach the bare surface as close as the passivated surface,  which  may also contribute to  the difference between the SFL observed in both cases.

To summarize, the SFL seems to be pinned at the top of the valence band nearly unaltered  by the  passivation of  Ge(001). This is in agreement with photoelectron observations~\cite{ARPES1}.  We  remark  that  the  insensitivity  of the SFL  to passivation explains conductance measurements in the same way as in Ref.~\onlinecite{APL}. Indeed, the inversion layer is not destroyed by hydrogen adsorption.

\subsection{Surface state structure}
There are two low-temperature Ge(001) reconstructions: p(2$\times$2) and c(4$\times$2). In both cases the surface atoms reduce the number of unsaturated bonds by forming bonded pairs~\cite{Zandt}. These dimers form long rows. Two atoms in a pair are not equivalent which  is manifest in a tilted geometry. The rows consist of  alternately   tilted pairs. The two  reconstructions  result from two possible relative orientations of pairs in adjacent  rows. These surface arrangements  appear  nearly equivalent energetically and can be reversibly switched in STM experiments\cite{japonce}.\\
Both experimental and theoretical findings regarding the Ge(001) surface point to the following structure of surface bands~\cite{Demkov}. The lower surface band, called usually $D_{up}$, is located very close to the top of the bulk valence band and it is a surface resonance. The name $D_{up}$ points to the fact that the related electron density  is concentrated on the upper dimer atoms. This band splits into two branches when going from $\bar{\Gamma}$ to $\bar{J}_{2}$ for either reconstruction. 
The electric neutrality of the surface requires  that this band be fully occupied.  There is an upper surface band  about 0.3-0.45~eV above the lower band. It is termed $D_{down}$ due to  maximal electronic density concentrated on the lower atoms in the dimers. 
This band stays in the bulk band-gap, which for germanium reads 0.742 eV at 0 K and 0.661 eV at room temperature. We note that in various studies the SFL always appears  between these two surface bands at variable  positions. \\
The data shown in Fig.~3 correlate well with the above description. The surface vs. bulk character of  the occupied states just below the SFL cannot be judged from our experiments. However, about 0.4~eV above SFL there is a peak  that we interpret as the  surface band since it is located within the bulk band gap at a position where a surface band is found in the  ARPES experiments~\cite{ARPES1,ARPES2} and  is  washed away upon passivation.\\
The only feature appearing in STS data we cannot firmly  identify  is a very weak and flat peak observed just above the SFL for Ge(001)-c($4\times 2$). It is worth noting that one can find such a structure in the STS data published by Gurlu~\cite{Gurlu}. It seems unlikely  to be  a surface band. It appears right above the valence band, where shallow acceptor states are located. The shallow elemental acceptors have  an  energy about 10-15 meV above the edge of the valence band, two vacancy ionization energies are located 0.05~eV and 0.14~eV above the valence band\cite{defect} and for the  vacancy-Sb complex transition states are located between 0.11-0.18 above the top of the valence band~\cite{Bracht}. These energies correlate with the peak width. However, it is not clear why those states are evident for this particular surface reconstruction only. 

Hydrogen termination of the Ge(001) surface does not alter bonds between the surface Ge atoms and, consequently, does not touch the rows on the surface. One hydrogen atom binds to every surface Ge atom turning the dimer atoms equivalent. The hydrogen terminated Ge(001) surface has not attracted as much attention as the clear surface, but there are several studies available such as both ARPES~\cite{ARPES1} results and recently published STS data~\cite{Avenues,Kolmer}. They point  out  similar surface state structure as in Fig.~3. The surface band gap increases to about 0.85 eV. ARPES data show  clearly  that the occupied surface states are located deeply below the top of the valence band (about 3-4 eV). There is no additional surface-related feature close to the Fermi level seen in the ARPES experiments. The unoccupied surface bands are above the lower bulk conductance bands~\cite{Kolmer}. 

The above resume  shows that the spectroscopic experiments explore the actual structure of the surface states. The absence of any surface state for Ge(001):H within the bulk band-gap makes the observed band bending a puzzle.  Noteworthy, the results obtained in the four point probe experiments  at Ge(001) and Ge(001):H show that dangling bonds (or any surface channels) do not considerably contribute to the electron transport even at sub-micron scale.

\subsection{Fermi level pinning}
As already mentioned, band bending at surfaces is due to the presence of the surface states not aligned with the bulk bands. In general, the bulk position of the Fermi level  leads to high occupation of the surface states  and, consequently, the surface gets charged. As a result,  an  electrostatic field appears changing the position of the surface Fermi level. In turn, the positions of bands are shifted with respect to the Fermi level at the surface and the surface charge is significantly reduced.\\
Having said this, it is evident that there is no reason for the Fermi level to appear at the top of the valence band on Ge(001). For the p-type doped sample, this could happen for  high enough doping. But for the n-type samples, the bulk Fermi level must be above the mid-gap.  At T=5~K the Fermi-Dirac factor excludes any significant thermal occupation of states located E=13~meV away from it, as
\begin{displaymath}
 \left[\exp\left(\dfrac{E}{k_{B}T}\right)+1\right]^{-1} =7.4\cdot 10^{-14}
\end{displaymath}
while the surface band density is of the order of $6\cdot 10^{14}$ states per cm$^{2}$. Hence, any occupation of the empty surface band on Ge(001) can be excluded. Furthermore, for SFL positioned 0.2 eV higher this occupation would not be significantly  different.   In this context, we  analyze three possibilities. First, for the neutral surface the mechanism behind  sticking of the SFL  to the top of the valence band is not clear. Indeed,  the theory of FLP gives rise to the following formula  (Schottky approximation~\cite{Monch}) relating the surface charge density $\delta_{s}$ needed for band bending $\phi$ and bulk dopant density $N_{b}$ 
\begin{equation}\label{charge}
\delta_{s}\sim \sqrt{\phi N_{b}}.
\end{equation}
The formula is valid within the band gap. For the neutral surface $\delta_{s}=0$ the band bending also vanishes.  But  experimentally it yields at least  $\phi\approx 0.3-0.4$~eV.
Second, one can think of superficial surface states located below the top of the valence band that, when occupied,  make the surface negatively charged. But there are good reasons to reject  the  presence of such states as neither theory nor experiment  hint at such a  possibility. These states should be present both on the bare and passivated surface. However  no surface states are observed in ARPES~\cite{ARPES1} of the hydrogen terminated surface. 
Third, the surface can be positively charged due to the states from the valence band above the SFL. But  positive charges do bend bands downwards to move the SFL away from the valence band. In the case of n-type doped germanium upward shifting of the bands is evident.  \\ 
Using eq.~(\ref{charge}) we  estimate that for the intrinsic slab the charge density needed to cause band bending of the order of 0.3-0.4~eV yields $5\cdot 10^{9}$~cm$^{-2}$, and  for the doped sample  this is $4 \cdot 10^{11}$~cm$^{-2}$. These estimates are  valid unless the SFL splits the valence band. Otherwise  much more charge is needed. In the first case, the density is small and one could  expect  that some surface irregularities can do this job. In the latter  case   the density is a substantial fraction of the defect density. We would probably notice some difference in the course of our experiments where defects appear neutral.

\subsection{Postulated scenario }
The above considerations strongly  suggest  that the FLP in germanium cannot be the conventional electronic  effect adjusting the SFL to the surface state structure. A  similar tendency has been observed for the silicon surfaces~\cite{si1, si2}, but due to  a  larger band  gap  the effects are not that spectacular.\\ 
It is concluded from density functional theory calculations in Ref.~\onlinecite{defect} that the Ge(001) surface  strongly attracts electrons, effectively pushing the SFL towards the valence band. In turn, the defects differently behave in a near surface region compared to the bulk.  This  surface feature seems to originate from  a higher amplitude of valence band wavefunctions close to the surface  rather  than from excessive additional surface states. In this case  the SFL should somehow deviate upon passivation. \\
The solution of   this puzzle that  we propose is inspired by putting two papers --- Ref.~\onlinecite{ARPES3} and Ref.~\onlinecite{Wolkow} --- together. The first one reports ARPES experiments at higher temperatures on germanium and silicon samples.  It is  observed  that Si(001) performs similar to the Ge(001) for a substantial charge accumulated in the surface states above the Fermi level without significant change of the  SFL position.   Note  that the silicon sample under investigation was flashed up to 1520~K. The latter article by J. L. Pitters  shows  that at such temperatures a depletion of dopant density in the subsurface layer is formed. For samples flashed at lower temperatures (1250~K) no migration of dopants is seen. A gradient in dopant density is evident for the samples treated  at  higher temperatures. As a consequence, the surface Fermi level is not entirely determined by its bulk value and the surface state structure, but also by  the dopant density at the surface. The only discrepancy in this comparison is that the sample investigated in Ref.~\onlinecite{ARPES3} is phosphorous doped while in the latter case it is  arsenic doped. \\
The usual approach to the SFL pinning holds that the distribution of dopants is  constant within the sample. The  only way the system can respond to perturbations (e.g. creation of a surface) is through electronic degrees of freedom. As such, the system stays in a constrained, but  not necessarily global, thermodynamic equilibrium. At  high enough temperatures,  where the diffusion is sufficiently quick  to adjust  the  dopant gradient to the perturbation, the true thermodynamic optimum is available.  When  cooling the  sample  the dopant gradient gets frozen.  If  the  cooling is quick enough  then  the gradient of dopants reflects the equilibrium Fermi level position at the temperature where the diffusion stops. Such  a  gradient profile  is not suited for lower temperatures making the SFL  at lower temperatures deviate from its equilibrium position. If it does not  entail  charging of the surface (the case close to our samples)  then no electrostatic field restoring the equilibrium SFL is created.\\
Within the one-electron approximation, the Fermi level (chemical potential) in semiconductors is calculated by imposing electric neutrality condition~\cite{Ibach}, i.e. the density of ionized donors $p_{d}N_{d}$ and holes $n_{h}$ has to be equal to the density of ionized acceptors $p_{a} N_{a}$ and electrons $n_{e}$ in conduction bands,
\begin{equation}\label{neutr}
p_{d}N_{d}+n_{h}=p_{a}N_{a}+ n_{e},
\end{equation}
where $p_{d}$ and $p_{a}$ denote the relevant ionization probabilities and $N_{d}$ and $N_{a}$ the dopant densities. Both the probabilities $\{p_{d},\  p_{a}\}$ and densities $\{n_{h}$, \   $n_{e}\}$ depend on the Fermi level via statistical factors. As such, eq.~(\ref{neutr}) can be used to set the Fermi level position. Now,  consider a slowly varying dopant density $N_{d}(\mathbf{x})$, $N_{a}(\mathbf{x})$. Then, we can locally, i.e. for every $\mathbf{x}$, solve the equation  
  \begin{equation}
p_{d}N_{d}(\mathbf{x})+n_{h}=p_{a}N_{a}(\mathbf{x})+ n_{e},
\end{equation}
to arrive at the Fermi level as a position dependent function. As such,  local dopant variation   impacts the electron Fermi level. For abrupt change of the dopant density  different calculation schemes are effective~\cite{phos}.

Finally we note, that in germanium    the ions are mobile enough  to ensure swift diffusion already at  800 K~\cite{Koffel,Bracht}. This makes our hypothesis natural as we heat the sample to about 1000~K. Furthermore, the ion sputtering we use is prone to produce a considerable amount of defects which may be stabilized by the electrostatic field and change the effective doping level, see Ref.~\onlinecite{Bracht}.\\

\section{Conclusion}
In this paper we described  electric transport experiments on the Ge(001) and Ge(001):H surfaces  both in the classical and tunneling regimes. They  show that the SFL is independent of the structure of surface states.  These results cannot be understood within the conventional mechanism behind FLP. We  argue that in our case a depletion of n-type dopants could be responsible for the experimental outcomes.

\acknowledgements
The authors wish to thank Jerzy Konior for instructive discussions.
This research was supported by the 7th Framework Program of the European Union Collaborative Project ICT (Information and Communication Technologies) "Planar Atomic and Molecular Scale Devices" (PAMS), Contract No. FP7-610446. M.K. acknowledges financial support received from the Foundation for Polish Science (FNP).
The research was carried out with the equipment purchased with the financial support of the European Regional Development Fund in the framework of the Polish Innovation Economy Operational Program (contract no. POIG.02.01.00-12-023/08).

\end{document}